\documentclass[
]{ceurart}

\usepackage{listings}
\newcommand\sbullet[1][.75]{\mathbin{\vcenter{\hbox{\scalebox{#1}{$\bullet$}}}}}

\DeclareMathOperator{\compose}{\sbullet}

\begin{document}

\copyrightyear{2022}
\copyrightclause{Copyright for this paper by its authors.
  Use permitted under Creative Commons License Attribution 4.0
  International (CC BY 4.0).}

\conference{Submitted to the PoEM 2026 Forum as part of the 19th IFIP WG 8.1 Working Conference on the Practice of Enterprise Modeling (PoEM 2026), December 1-3, 2026, Mytilene, Lesvos, Greece}

\title{From Trust-by-Design to Continuous Trust in the Era of Artificial Intelligence}

\author[1,2]{Peter Fettke}[%
orcid=0000-0002-0624-4431,
email=peter.fettke@dfki.de,
]
\cormark[1]
\address[1]{German Research Center for Artificial Intelligence (DFKI), Saarbr\"ucken, Germany}
\address[2]{Saarland University, Saarbr\"ucken, Germany}

\author[3]{Wolfgang Reisig}[%
orcid=0000-0002-7026-2810,
email=reisig@informatik.hu-berlin.de,
]
\address[3]{Humboldt-Universität zu Berlin, Berlin, Germany}

\cortext[1]{Corresponding author.}

\begin{abstract}
Trust is crucial in the digital world, particularly in the age of systems based on artificial intelligence. Traditionally, system correctness is demonstrated by proving that the system satisfies the required properties. However, digital systems involve humans, who can make mistakes, break rules, or even intentionally cheat the system. Technical devices may also malfunction or cause errors. Therefore, external stakeholders need to trust a system throughout its operation. To address this challenge, we propose the concept of \emph{continuous trust}. In this paper, we present a detailed explanation of this concept, based on a running example of a business’s refund department. The paper concludes with a discussion of related work, such as compliance checking and certification, as well as an outlook on further questions regarding the formalization and the necessary tool support.
\end{abstract}


\maketitle

\section{Introduction: The problem of continuous trust}
Classically, \emph{correctness} is considered within the framework of a software program together with a formal specification of the program's required properties. Formal methods are used for \emph{verification}, i.e., to prove that the software fulfills these properties. For reactive systems, \emph{temporal logic} is the standard technique nowadays: properties are represented as logical expressions that are to hold in single states or processes, i.e., sequences of states. An external stakeholder then must trust that automated proofs are correct. This is \emph{trust-by-design}.

Systems in the digital world, in particular enterprises, include not only software but also human activities and technical devices. Trust by design fails for systems in the digital world: a mistake by an involved human, or a malfunction of a technical device, may cause incorrect processes, even if the software is verified. Even more, in the era of artificial intelligence (AI), such systems can emerge autonomously, with software no longer under any human control. Finally, an outside stakeholder may want to trust a process’s correctness as it evolves; for instance, a compliance violation may not be detectable later. This is \emph{continuous trust}. Hence, this paper asks: how can continuous trust be organized?

The rest of this paper is divided into four sections. Section~\ref{sec:2} discusses the relationship between the world, a model of the world, and observed processes. Section~\ref{sec:3} is the main part of the paper and presents a running example to illustrate the concept of continuous trust. Section~\ref{sec:4} discusses the idea and its relation to prior work. Section~\ref{sec:5} concludes the paper with some general remarks.

\section{Integrating events, observations, and models}\label{sec:2}

To organize continuous trust for a system, the system's processes must be visible to outside stakeholders. A solution to this problem: during process evolution, the system provides information, e.g. logs or other data, and promptly communicates it to external stakeholders who observe the system’s processes. The stakeholders should be able to quickly assess whether the provided data correspond to one of the intended processes. This requires a clear description of the intended processes. To this end, a \emph{model} of the desired processes is necessary. Hence, three artifacts must work together:

\begin{itemize}

\item Events that happen during an evolving process of the system in \emph{the world}; 

\item Records of these events that are observable from the outside, or documented by the company in “logs”. In the sequel, we denote such records and logs as \emph{observations};

\item A formal \emph{model} that describes the desired processes.

\end{itemize}

\begin{figure}
   \centering
   \includegraphics[scale=.8]{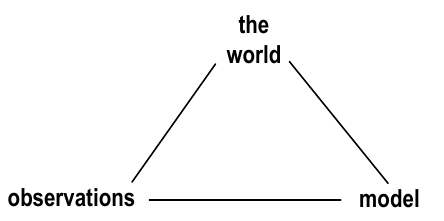}
   \caption{The relationship between the world, observations, and models}
   \label{fig:1}
\end{figure}

Figure~\ref{fig:1} sketches those artifacts.

The question is how these three artifacts integrate, how mistakes and inconsistencies that occur during process evolution are quickly detected, and how they together guarantee continuous trust. 

To this end we observe that each of the three artifacts generates its specific kind of processes:

\begin{itemize}

\item processes as they evolve in the world;

\item processes composed from observations; 

\item processes specified in a model.

\end{itemize}

Ideally, the processes of all three types would closely resemble each other. In mathematical terms, this would mean that they are \emph{isomorphic}. However, we can’t guarantee that a running process in the modeled system perfectly matches a process specified by the model. For example, a process of an enterprise may violate a compliance rule, even though the enterprise model properly reflects that rule. Worse, nobody may observe this violation. 

\begin{figure}
   \centering
   \includegraphics[scale=.8]{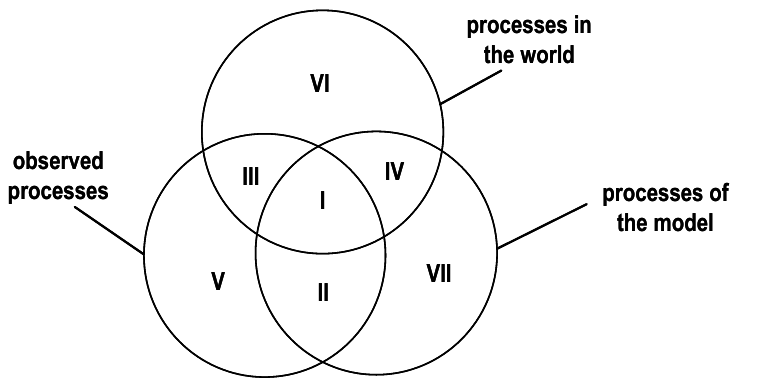}
   \caption{Classification of processes}
   \label{fig:2}
\end{figure}

Technically, this means that a process of one kind has no isomorphic counterpart in one or the other two types. This results in seven groups of processes. Figure~\ref{fig:2} sketches the different possibilities. Specifically, for a process $R$, together with its isomorphic counterparts, the following cases are distinguishable:

\begin{itemize}

\item $R$ in I: This is the ideal case: $R$ is part of the system’s behavior, has been observed, and is represented in the model. 

\item $R$ in II: $R$ is not part of the system’s behavior, yet $R$ has been observed and is represented in the model. This happens when the observations are flawed, and the model is generated from them.

\item $R$ in III: $R$ is not represented in the model, but it is part of the world and was also observed. This happens when the model does not accurately represent the world.

\item $R$ in IV: $R$ was not observed, but is part of the world and is represented in the model. This occurs when, in the ongoing process, an event of $R$ has not yet occurred.

\item $R$ in V: $R$ was observed, but is not part of the world and is not represented in the model. This happens when the system behavior is observed incorrectly.

\item $R$ in VI: $R$ is part of the world, but it was not observed and is not represented in the model. This happens when $R$ is unintended.

\item $R$ in VII: $R$ is represented in the model, but it is not part of the world and was not observed. This happens when $R$ is not really important.

\end{itemize}

In summary, the processes in category II–VII point to a fault in the overall framework of Figure~\ref{fig:1}. Continuous trust is realized when the observed processes do not differ from those in the real world.

\section{Running example: A company's refund department}\label{sec:3}

The following real-world example shows that the notion of continuous trust is in fact useful to better understand a system and to trust in its correctness. This includes the classification of processes as in Figure~\ref{fig:2}.

\subsection{The setting}\label{sec:3:1}

Here we consider the example of a retail company's refund procedure. This example is inspired by our prior work on a case study in \cite{fettke2026trust}.

The company’s refund department handles customer refund requests, so-called \emph{claims}. Each claim involves a \emph{customer} and an \emph{item}, and is handled by a company \emph{staff} member, who either \emph{approves} or \emph{rejects} it. This is accompanied by several constraints:

\begin{enumerate}

\item Specialized staff: For each article $a$, there is a designated set of staff, $f(a)$, authorized to handle claims related to $a$. Every incoming claim for $a$ is assigned to a staff member from the set $f(a)$.

\item Conflict of interest: Staff members may also be customers of the retail company. It must be strictly ensured that no staff member processes their own claim.

\item Limited replacement: The assigned staff member may be replaced by any other authorized staff member. For each claim, this may happen at most once.

\item Proper termination: Each customer claim is ultimately either accepted or rejected.  

\end{enumerate}

In our previous work \cite{fettke2026trust}, we already demonstrated how to prove these four constraints. In the following, we present a systematic approach to determining whether a company adheres to these requirements when carrying out a process.

\subsection{Statics in the lifeworld: structures}\label{sec:3:2}

\begin{figure}
   \centering
   \includegraphics[scale=.9]{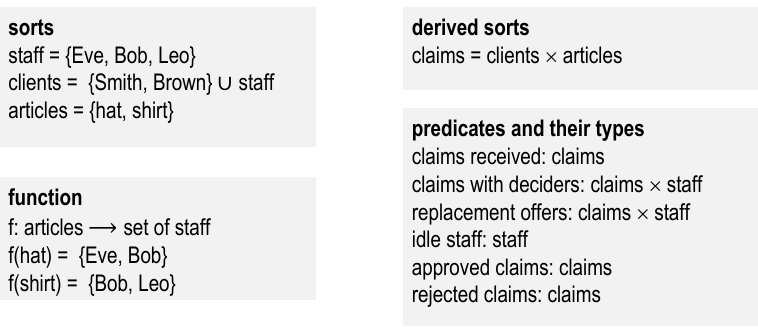}
   \caption{Sorts, functions, and predicates of a given refund procedure}
   \label{fig:3}
\end{figure}

To understand a system -- particularly a company -- static facts and the relationships between them are organized into an \emph{algebraic structure}. Such a structure consists of a finite number of sets, so-called \emph{basic sorts} of elementary objects that occur within the company. Complex objects can be composed from these, so-called \emph{derived sorts}. Functions and predicates are identified based on these sorts. Figure~\ref{fig:3} shows, as an example, the algebraic structure of a refund department from Section~\ref{sec:3:1}.

The concept of algebraic structures has many advantages: It is sufficiently general to provide an initial formal framework for key aspects of a company. Key properties of companies can be appropriately formulated using predicates over the basic sorts. Sorts, functions, and predicates are not symbolic objects here, but rather items of the real world. They motivate symbolic representations such as those used in predicate logic.

Note, we restrict the usual concept of an algebraic structure by assuming only finite basic sorts. This is sufficient for now; it allows every element to be represented symbolically as a constant. Questions regarding the representability of elements using ground terms then do not arise.

\subsection{Dynamics in the lifeworld: predicate updates}\label{sec:3:3}

As explained in Section~\ref{sec:3:2}, algebraic structures represent unchanging facts and relationships within a system -- in the current example, a refund department. Here, we will now focus on \emph{dynamic changes}. Dynamic changes consist of individual steps. Typical examples are the two steps:

\begin{enumerate}

\item \emph{Receive claim}: The refund department receives a claim from customer \emph{Brown} regarding the item hat.

\item \emph{Select decider}: The refund department selects staff member \emph{Eve} to handle the claim.

\end{enumerate}

In the context of algebraic structures, a step refers to the update of one or more predicates. In the current example, the first step updates the predicate “claims received”. As a result of this step, “claims received” now holds for the tuple \emph{(Brown, hat)}. The second step can occur if “claims received” holds for \emph{(Brown, hat)} and “idle staff” holds for \emph{Eve}. As a result of this step, two predicates, “claim with deciders” and “replacement offer”, both hold for the triple \emph{(Brown, hat, Eve)}.

\begin{figure}
   \centering
   \includegraphics[width=1\textwidth]{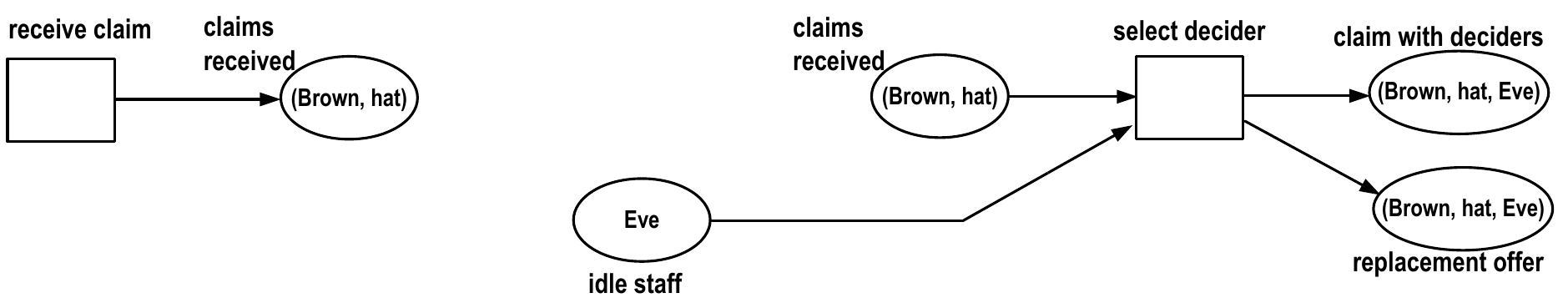}
   \caption{Two steps}
   \label{fig:4ab:4}
\end{figure}

Figure~\ref{fig:4ab:4} shows these steps graphically: a predicate $p$ is represented as an ellipse; an element $a$ lies within the ellipse if $p$ holds for $a$. A step is represented as a rectangle. An arrow indicates the effect of a step $s$ on $p$: The arrow ends at $p$ if, as a result of $s$, an element satisfies the predicate $p$. An arrow begins at $p$ if $p$ no longer holds after the step. Figures~\ref{fig:4ab:4} show the two steps described before.

\begin{figure}
   \centering
   \includegraphics[scale=.45]{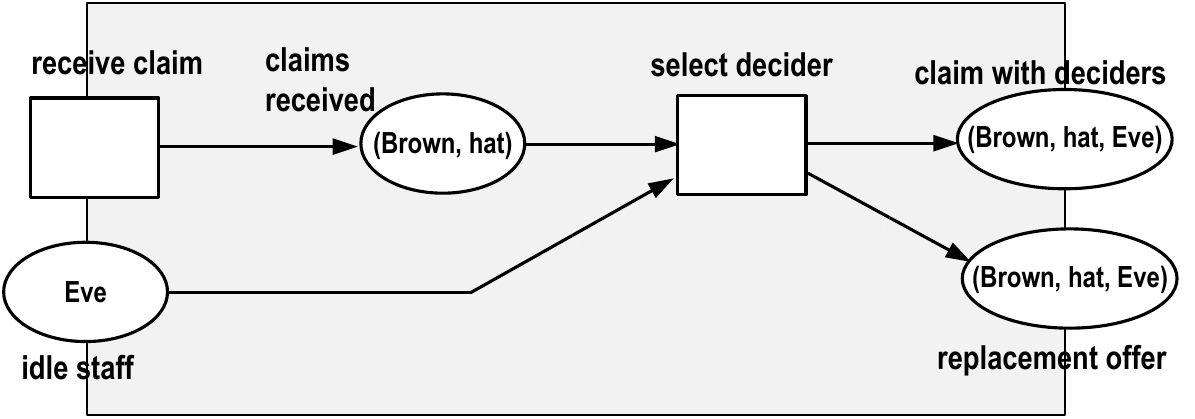}
   \caption{The process \emph{receive claim} $\compose$ \emph{select decoder}, composed of the steps of Figure~\ref{fig:4ab:4}}
   \label{fig:4c:5}
\end{figure}

\begin{figure}
   \centering
   \includegraphics[width=1\textwidth]{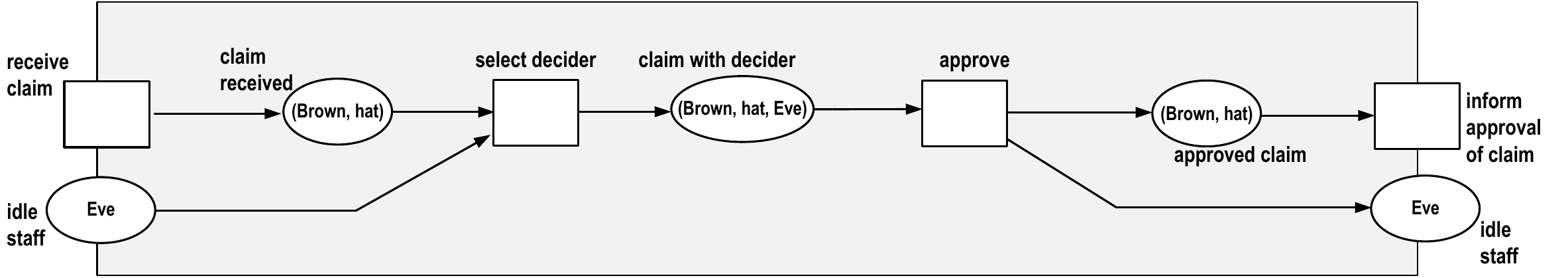}
   \caption{Process $r_1$}
   \label{fig:5a:6}
\end{figure}

\begin{figure}
   \centering
   \includegraphics[width=1\textwidth]{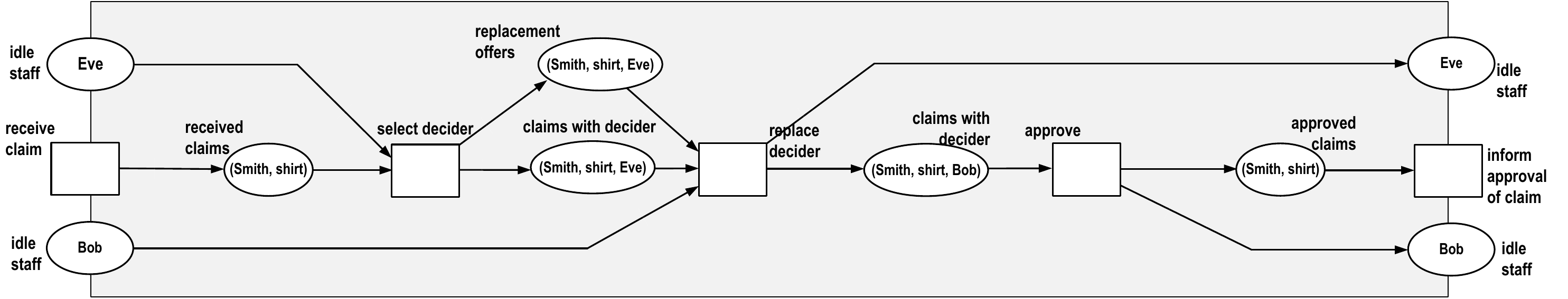}
   \caption{Process $r_3$}
   \label{fig:5c:8}
\end{figure}

Steps can now be combined to form sequences and more complex behavior, called “processes”. Figure~\ref{fig:4c:5} shows the composition of the steps from Figure~\ref{fig:4ab:4}. Finally, Figure~\ref{fig:5c:8} extends the behavior shown in Figure~\ref{fig:4c:5} by adding two additional predicates and two additional steps. Figure~\ref{fig:5a:6} illustrates a typical intended behavior of the refund department. Technically, this simply describes how certain objects enter and exit the scope of specific predicates.

\begin{figure}
   \centering
   \includegraphics[width=1\textwidth]{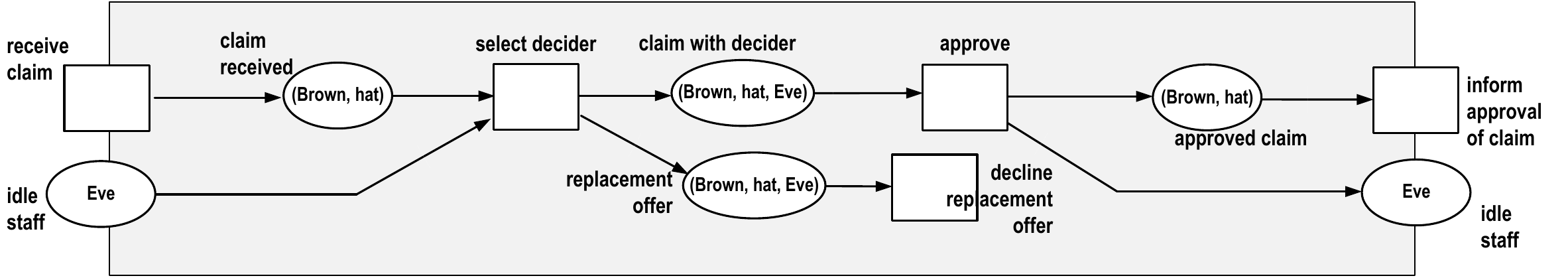}
   \caption{Process $r_2$}
   \label{fig:5b:7}
\end{figure}

As a variation, Figure~\ref{fig:5b:7} extends the process shown in Figure~\ref{fig:5a:6} by adding the option to replace the decider \emph{Eve} with another decider. However, this offer is rejected. Figure~\ref{fig:5c:8} shows another variation: First, the staff member \emph{Bob} is named as the decider, but is then replaced by \emph{Eve}.

\subsection{A model of the retail company's refund procedure}

Next, we present a model of the retail company’s refund procedure. Finite structures in the real world, such as the one shown in Figure~\ref{fig:3}, can be represented symbolically and formally with complete precision: there is a symbol for every element; for every function $f$, the pairs \emph{(a, f(a))} are specified explicitly; and for every predicate $p$, the elements to which $p$ applies are listed. Things get more complicated with infinite structures, but we’ll leave that aside for now.

\begin{figure}
   \centering
   \includegraphics[width=1\textwidth]{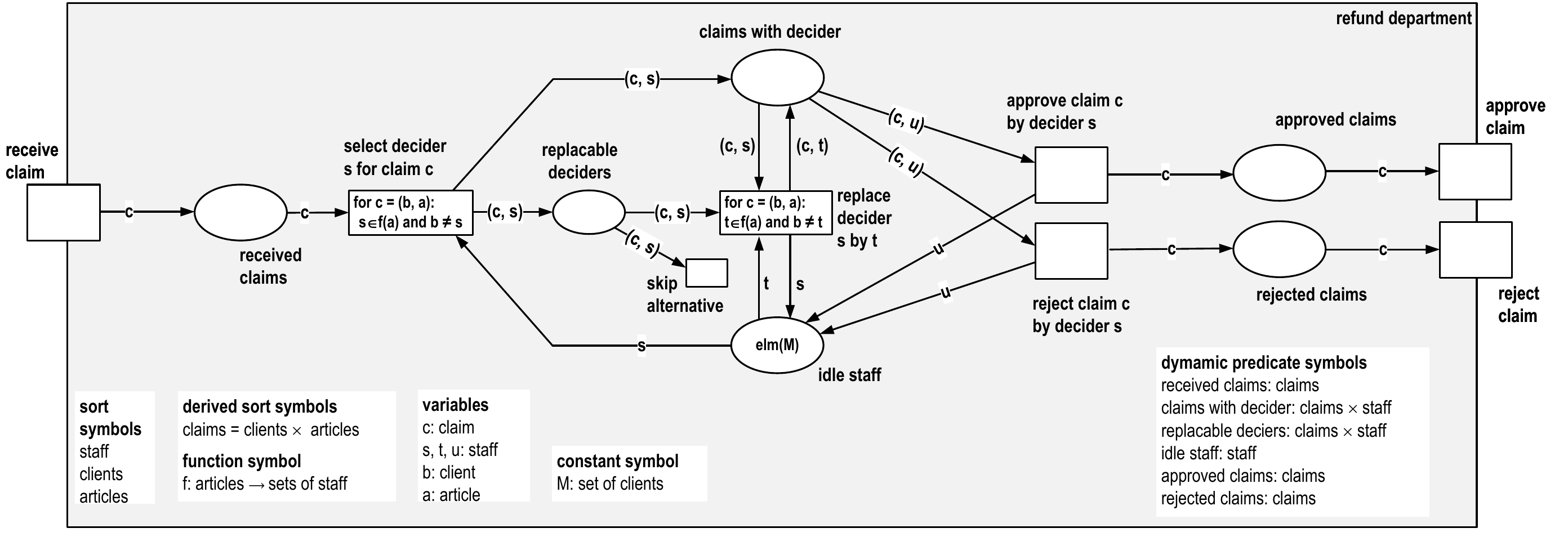}
   \caption{Behavior of the refund department}
   \label{fig:6:9}
\end{figure}

Figure~\ref{fig:6:9} shows a graphical representation of the refund department. Here, steps are parameterized using \emph{variables} on the arrows. A step according to Section~\ref{sec:3:3} is created by instantiating the variables with constant elements. Figure~\ref{fig:6:9} includes the step “receive claim” where $c$ is a variable for claims. Instantiating $c$ with \emph{(Brown, hat)} yields the step in Figure~\ref{fig:4ab:4}. Likewise, Figure~\ref{fig:6:9} represents the steps in Figure~\ref{fig:4ab:4} and the processes in Figure~\ref{fig:5a:6}. For technical details of this kind of system representation and its processes, we refer to \cite{Fettke_Reisig_24}.

The concept of updating predicates yields models with a most tight connection to the world. All processes show objects entering and leaving predicates. Other concepts, such as assignment statements to program variables, are not used. By limiting ourselves to variables as we know them from mathematics and logic, these processes involve only predicates and their updates -- concepts that reflect the reality of the refund department’s operations.

\subsection{Unintended processes}

The notion of processes of a system, as discussed in Section~\ref{sec:3:3}, includes many details of the system. Therefore, any mistake, i.e., any process that violates one of the constraints in Section~\ref{sec:3:1}, belongs to one of the classes II – VII of Figure~\ref{fig:2}, and thus can be detected by an observer.

\begin{figure}
   \centering
   \includegraphics[width=1\textwidth]{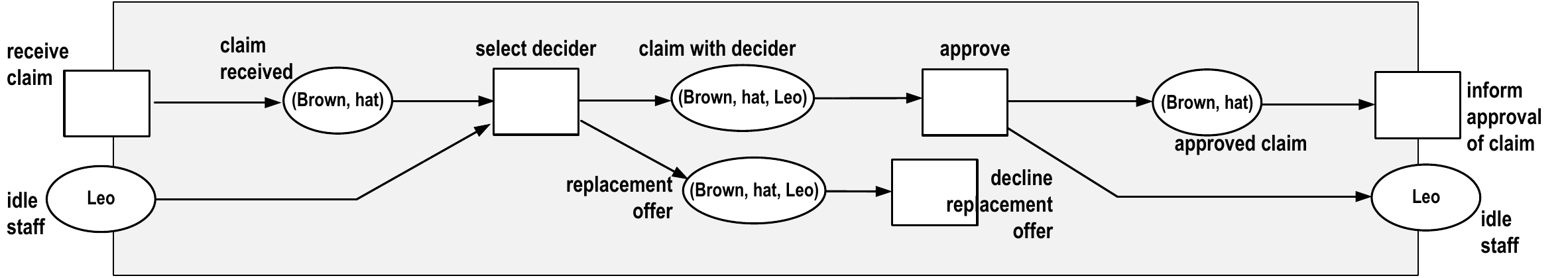}
   \caption{Unintended process: staff member \emph{Leo} is not authorized to handle a claim for the article hat.}
   \label{fig:7a:10}
\end{figure}

An interesting side note: Although an observer typically observes an order between any two events, the resulting process, constructed by composing all observed process steps, typically orders events only partially. For example, the process of Figure~\ref{fig:7a:10} can be constructed by observing the process steps: \emph{receive claim}, \emph{select decider}, \emph{approve}, \emph{decline replacement offer}, and \emph{inform approval of claim}, resulting in the composition:

\begin{quote}
\emph{receive claim} $\compose$ \emph{select decider} $\compose$ \emph{approve} $\compose$ \emph{decline replacement offer} $\compose$ \emph{inform approval of claim}.
\end{quote}

But the event \emph{decline replacement} offer is not ordered with \emph{approve} or with \emph{inform approval of claim}. 

In the following Figures~\ref{fig:7a:10} to \ref{fig:9b:16}, we show some typical examples of this kind of unintended processes. The processes described are not processes of the model for various reasons:

\begin{itemize}

\item Figures~\ref{fig:7a:10} to \ref{fig:7c:12}: Three processes with unauthorized deciders.

\item Figure~\ref{fig:8:13}: There is a conflict of interest since \emph{Bob} is processing his own claim.

\item Figure~\ref{fig:8:14}: The assignment of \emph{Eve} is correct. But her replacement by \emph{Bob} is not.

\item Figure~\ref{fig:9a:15}: The person responsible for deciding the claim has changed more than once.

\item Figure~\ref{fig:9b:16}: This claim is never handled by a decider.

\end{itemize}

\begin{figure}
   \centering
   \includegraphics[width=1\textwidth]{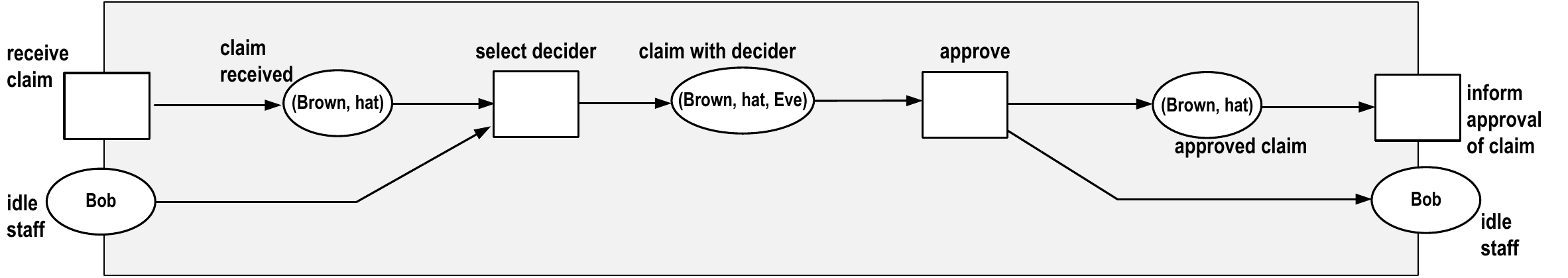}
   \caption{Unintended process: staff member \emph{Eve} is not authorized to handle a claim for the article hat without a replacement order}
   \label{fig:7b:11}
\end{figure}

\begin{figure}
   \centering
   \includegraphics[width=1\textwidth]{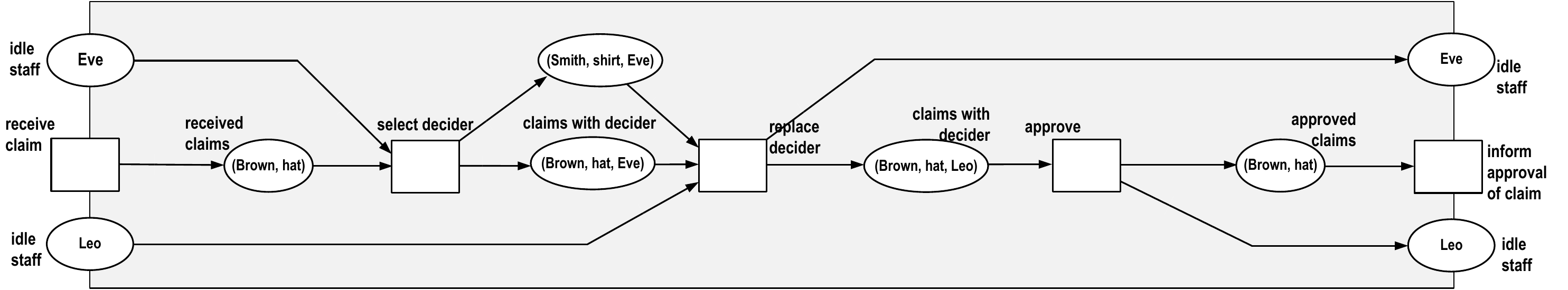}
   \caption{Unintended processes: staff member \emph{Leo} is not authorized to handle a claim for the article hat. Notice that the authorized decider \emph{Eve} is selected first, but is then replaced by \emph{Leo}.}
   \label{fig:7c:12}
\end{figure}

\begin{figure}
   \centering
   \includegraphics[width=1\textwidth]{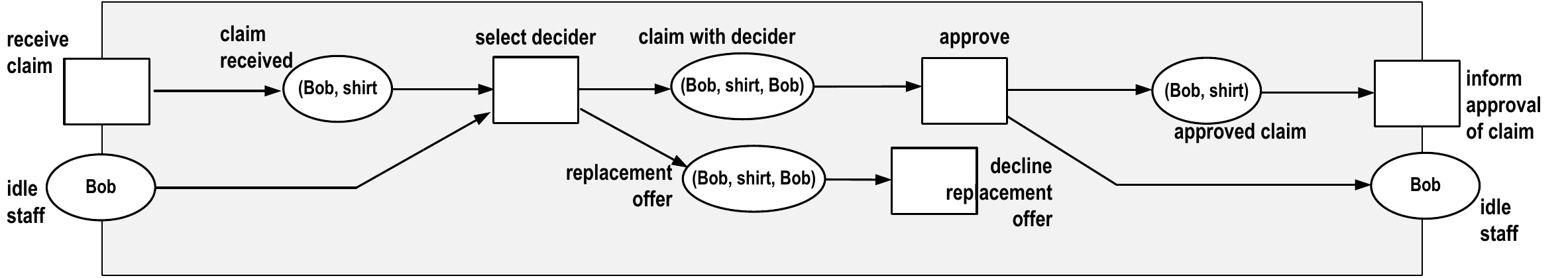}
   \caption{Unintended conflict of interest in the process: staff member processes his own claim}
   \label{fig:8:13}
\end{figure}

\begin{figure}
   \centering
   \includegraphics[width=1\textwidth]{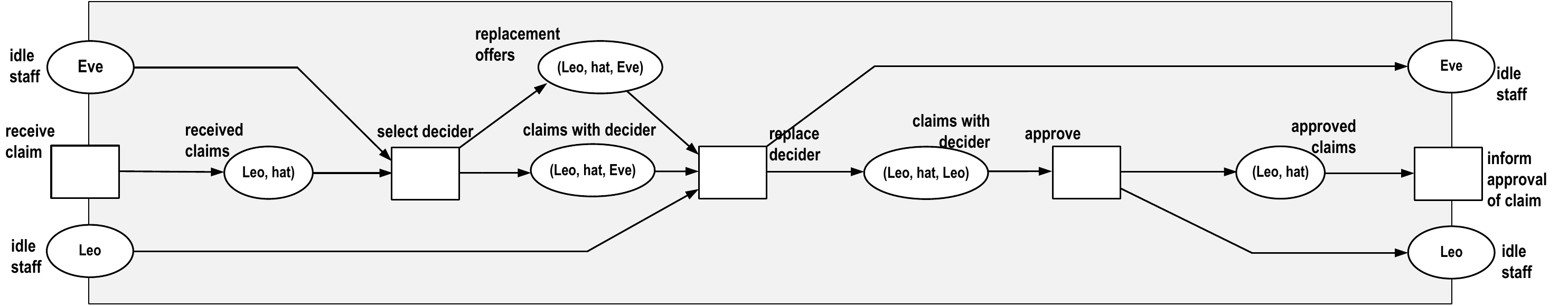}
   \caption{Unintended conflict of interest in the process: staff member processes his own claim. Note: first, he authorized decider \emph{Eve} is selected, but then replaced by \emph{Leo}.}
   \label{fig:8:14}
\end{figure}

\begin{figure}
   \centering
   \includegraphics[width=1\textwidth]{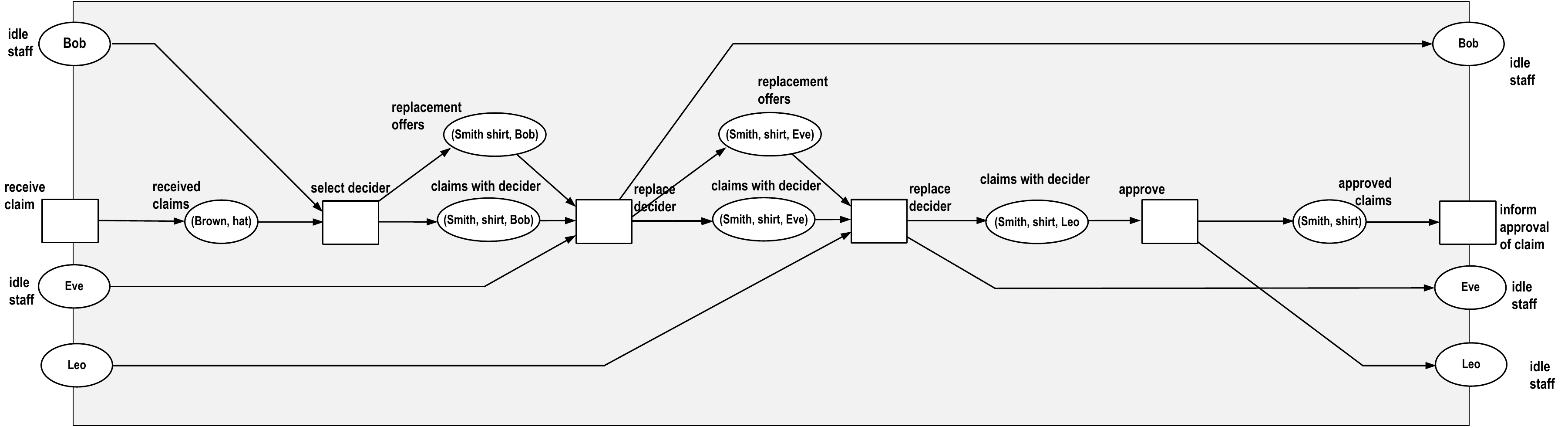}
   \caption{Unintended process: the decider to handle the claim is replaced more than once}
   \label{fig:9a:15}
\end{figure}

\begin{figure}
   \centering
   \includegraphics[scale=.45]{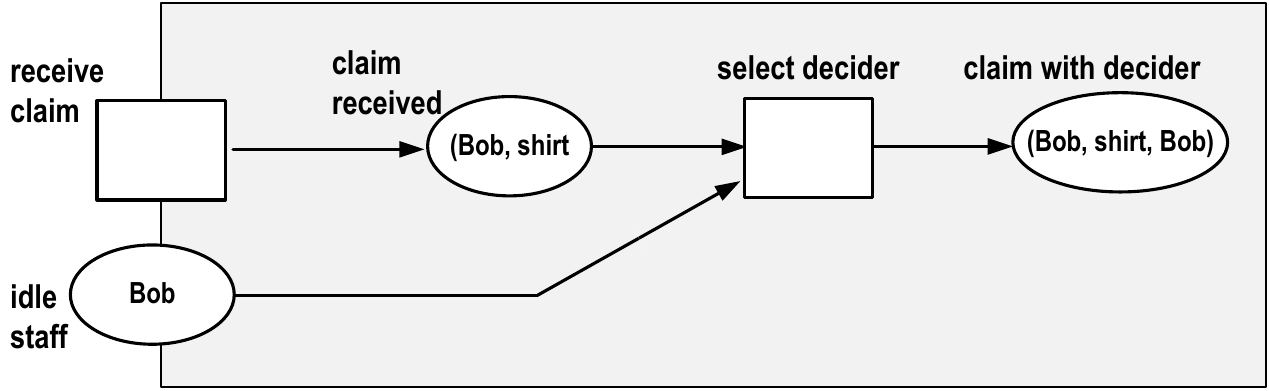}
   \caption{Unintended process since improper termination: a customer claim is neither accepted nor rejected}
   \label{fig:9b:16}
\end{figure}

An interesting question arises when we consider which category of processes in Figure~\ref{fig:2} the unintended processes fall into. By the assumption of this example, all unintended processes are observed processes. Furthermore, it has been demonstrated that none of these processes belong to the model's processes. Therefore, the processes cannot fall into categories I and II. This leaves categories III and V. If one argues that an unintended process falls into category V, this would mean the observed process is incorrect because the real process would differ from the observed one. Assuming the observed processes are correct, then all unintended processes fall into category III.

\section{Discussion and related work}\label{sec:4}

Edsger W. Dijkstra has repeatedly proposed to strictly distinguish separating the “correctness problem” and the “pleasantness problem” in computer science, and proposed a “firewall” between them \cite{dijkstra:firewallcacm}. His reasoning: The methods for solving the formal, mathematical “correctness problem” faced by computer scientists differ fundamentally from the methods for solving the informal “pleasantness problem” faced by users. In this context, a model is always limited to one side or the other of this wall.

Our case study shows that both sides of Dijkstra’s firewall can build upon the same foundations. This approach has many advantages and provides significant benefits: We can capture the core ideas of our natural, intuitive understanding of the world, enrich them with formal concepts and use the model to develop software systems. Therefore, we cannot only prove that a system model is correct; we can also prove that observations made during the system's evolution fit the system's rules. In other words, we can continuously trust the system.

Numerous studies from various research fields address the challenges in the context of trust in digital systems. Examples include “Reactive Systems” by Aceto et al. \cite{aceto2007reactive} from the perspective of software development and “Conformance Checking” by Carmona et al. \cite{carmona2018conformance} from the perspective of business process management. 

The above introduction of three sets of processes is also similar to the study of behavior introduced in \cite{Buijs2014}. However, there is an important difference: The authors of that work assume that behavior can be described in many forms, e.g. \emph{traces} or \emph{process algebras}. In fact, in that model, processes are often identified as formal words, i.e. \emph{strings}. In contrast, we propose to understand processes in the world, model, and logs as abstract concepts, namely as a \emph{partially} ordered steps; each step refers to an update of one or more predicates. This kind of process is sometimes called \emph{asynchronous} \cite{lamport2026science}. 

Trust is a topic of particular interest in the context of AI systems, e.g. \cite{HenriqueSantos2024,10.1145/3789256}. Our idea clearly contributes to the motivation behind these studies. However, our ideas are not restricted to software systems that use the latest AI innovations in machine learning, such as large language models.

The concept of software certification presented by Mehlhorn et al.  \cite{McConnellMehlhorn2011} is related to our ideas. For instance, an observed process can serve as a witness of specific system properties. While Mehlhorn et al. focus specifically on software certification, our focus is on studying processes in the digital world.

\section{Conclusion}\label{sec:5}

Demonstrating relevant properties is central to maintaining trust in systems. To this end, we have proposed the concept of \emph{continuous trust}, which ensures trust in a system even during operation. Rather than proposing a new specification language, we have used elementary concepts from mathematical set theory and logic.

This yields models that closely approximate reality. We illustrated these concepts using the example of a refund department. Future research must further develop the presented framework. For example, continuous trust is based on the idea that observations evolve over time and are never complete. In other words, tools for continuous trust must check trust continuously, not just once. Over time, more evidence can be acquired to prove or disprove specific system properties.

Furthermore, it is evident that suitable tools are necessary for practical applications. These tools will provide hands-on experience with the presented ideas and concepts in various application domains.


\section*{Declaration on Generative AI}
  The authors have not employed any Generative AI tools.

\bibliography{main}

\appendix

\end{document}